\documentclass[%
 reprint,
 amsmath,amssymb,
 aps,
]{revtex4-2}

\usepackage{footmisc}
\usepackage{hyperref}% add hypertext capabilities
\usepackage[mathlines]{lineno}% Enable numbering of text and display math
\usepackage{xcolor}
\usepackage{soul}

\usepackage{comment}
\usepackage{mathtools}

\usepackage{graphicx}% Include figure files
\usepackage{dcolumn}% Align table columns on decimal point
\usepackage{bm}% bold math
\begin{document}

\preprint{APS/123-QED}

\title{Acoustically-Coupled MEMS Transducer Pairs with Loss and Gain}

\author{Samer Houri}
 \altaffiliation{imec, Heverlee, Belgium}%
 \email{Samer.Houri@imec.be}
 \author{Rachid Haouari}%
\author{Bart P. Weekers}%
\author{Veronique Rochus}%

%\affiliation{
%\affiliation{
%\affiliation{%

\date{\today}% It is always \today, today,

\begin{abstract}
This work treats the dynamics of pairs of microelectromechanical ultrasound transducers (MUTs) that are immersed in water and acoustically coupled through the fluid medium. A series of these transducer pairs with varying diameters (and thus resonance frequency) and pitch separation (and thus coupling strength) are fabricated and measured. The work presented here models and quantifies the open-loop coupling between the MEMS transducer pairs and its dependence on pitch. Furthermore, a gain feedback loop is systematically applied to one of the device pair and the dynamics of the acoustically-coupled gain-loss system is investigated, and the formation of an exceptional-point or of an Hopf bifurcation is equally used to quantify the coupling coefficient. This work provides an experimental study of acoustic coupling in MUT transducers, as well as an exploration of the formation of exceptional points in acoustically-coupled MEMS transducers.%This work provides an experimental study of the coupling in MUT transducers including near-field and elastic-coupling effects, as well as offering an analytical and an experimental exploration of the onset of strong coupling in coupled MEMS transducers and the formation of exceptional points.%With the gain feedback loop applied, the dynamics of the system are captured by a coupled oscillator-resonator pair, whose parameters depend on the coupling strength and the exact configuration of the feedback loop.

%\begin{description}
%\end{description}}

\end{abstract}

\maketitle

%\tableofcontents

%\section{\label{sec:level1}INTRODUCTION}

\indent\indent\ The effects of coupling on interacting oscillators and resonators have been investigated since Huygens first observed the sympathetic oscillations of suspended clocks in the eighteenth century \cite{huygens1895letters}. Further progress in dynamics helped distinguish between the impact of weak and strong coupling \cite{dolfo2018damping} in interacting systems, where weak coupling plays a significant role in the synchronization of self-sustained oscillators (simply referred to as oscillators from hereon) \cite{pikovsky2003synchronization}, whereas strong coupling, which is defined as a coupling rate that exceeds detuning and damping, is most relevant to passive resonators (referred to as resonators from hereon).\\
\indent\indent\ In Micro- and Nano-Electro-mechanical Systems (M/NEMS) examples of coupled resonators and oscillators abound, with various coupling mechanisms being demonstrated such as mechanical coupling due to under etch \cite{kenig2009intrinsic,marquez2017asymmetrically} or mechanical linkages \cite{shim2007synchronized} for instance, electrical coupling due to fringing fields \cite{krylov2014collective} or electrical gates \cite{siskins2021tunable}, and optical coupling \cite{piergentili2018two}. Indeed, linear and nonlinear coupling of physically distinct M/NEMS resonators have been leveraged to enable improved sensors \cite{chang2023serial,spletzer2006ultrasensitive,zhao2016review,qiao2023frequency}, as well as to enable basic research such as the observation of symmetry-breaking and exceptional points (EP) in systems of coupled M/NEMS devices \cite{zhang2024exceptional,zhou2023higher}. \\
\indent\indent\ One mechanism for coupling MEMS devices is via acoustic waves through a fluid medium as opposed to elastic coupling through the solid substrate on which the devices reside \cite{yao2020design,annessi2022innovative}. This mode of coupling has been recognized as playing an important role in the design of large arrays of MEMS transducers such as those found in Micromachined Ultrasound Transducers (MUTs) \cite{pritchard1960mutual,porter1964self,meynier2010multiscale,turnbull2002fabrication,maadi2016self,oguz2013equivalent,weekers2021design}. Despite it playing a significant role in the design of practical M/NEMS transducer arrays experimental studies, systematically exploring and quantifying acousto-fluidic coupling, remain scarce. This is further complicated by the fact that a systematic study needs to account for both, the unavoidably present elastic coupling \cite{vysotskyi2021comparative} as well as the eventual acoustic coupling in a fluid medium.\\% For simplicity, from hereon we refer to the coupling due to the acoustic waves in water simply as acoustic coupling, and for the coupling through the substrate simply as elastic coupling.\\
%\indent\indent\ Radiation resistance, nonlinearity, near field effects....\\
\indent\indent\ In this work, we systematically characterize the impact of coupling between pairs of equally sized circular Piezoelectric Micromachined Ultrasound Transducers (PMUTs) with varying pitches (center-to-center separation), this is done for varying diameters both in-air, where the acoustic coupling is considered negligible, and in-water where the acoustic coupling is dominant. In the open-loop regime, i.e., without the presence of a gain feedback loop, we are able to investigate the dependence of the strength and phase of the coupling on the pitch separation, while the application of a gain feedback loop leads to self-oscillation whose onset is dependent on the coupling's strength and phase. We thus leverage the emergence of self-oscillation as an additional means to experimentally quantify the coupling between the PMUT devices.\\
%\indent\indent\ The device pairs used in this work are shown in Fig.~1(a), one clearly visible aspect is the lack of electrical connection to one of the devices, this is because while both devices possess the same stack of layers [Silterra] and both form suspended membranes, one of the devices is a mechanical dummy, i.e., is not wired to undergo electrical excitation, this leads to a slight discrepancy in the resonance frequencies of both devices, however, the impact of coupling remains clearly visible.\\
%\section{Modeling}
\indent\indent\ The measured PMUT devices exhibit quality factors on the order of hundreds in air and of tens in water. Despite the low quality factor in water, it is still possible to use a single mode expansion to describe the dynamics of these MEMS devices \cite{liem2021acoustic}. Similarly, the dynamics of the two coupled PMUTs can be described by two coupled harmonic resonators. %A peculiarity of the studied system in this work is that two coupling components need to be included, these are elastic coupling through the substrate, and radiative coupling (i.e., acoustic coupling through the liquid).
Since, by design, only one of the two devices is subject to a transduction, we refer to this device from hereon as device 1 or PMUT1, wherein the forcing (or electronic feedback) terms are only applied to this device.\\
\indent\indent\ Thus, the coupled PMUTs are governed by the equations of coupled harmonic resonators with one of the resonators subject to forcing and feedback, these read\\
\begin{equation}
\begin{split}
\ddot{x}_1(t) + (\gamma_1 + \beta x_1^2)\dot{x}_1 - \Gamma_{ac}\dot{x}_2 + \omega_1^2x_1 + \alpha x_1^3 - \Gamma_{el}x_2 = \\
F(t) + g(\dot{x}_1,\dot{x}_2)\\%,\dot{x}_2)\\
\ddot{x}_2(t) + \gamma_2 \dot{x}_2 - \Gamma_{ac}\dot{x}_1 + \omega_2^2x_2 - \Gamma_{el}x_1 = 0\\
\end{split}
\label{eqn:Eq1}
\end{equation}\\
where $x_{1, 2}$, $\gamma_{1, 2}$, and $\omega_{1, 2}$, are the modal displacements, the linear dissipations, and the natural frequencies, of PMUT1 and PMUT2, respectively. The dot symbol indicates time derivative. Since only PMUT1 is ultimately subject to forcing and potentially to large amplitude vibrations, additional terms $\alpha$ and $\beta$, are added in the governing equation of device 1, where $\alpha$ and $\beta$ are the nonlinear Duffing and nonlinear damping experienced by PMUT1 under the influence of large amplitudes of vibration. $F(t) = F\cos{(\omega t)}$ is the direct harmonic forcing term where $\omega$ is the drive frequency \cite{houri2025comparing}, while $g(\dot{x}_1,\dot{x}_2)$ is the feedback forcing term which is set to zero in the case of open-loop operation. Finally, $\Gamma_{ac}$ and $\Gamma_{el}$ are the complex-valued acoustic and elastic coupling coefficients, respectively. \\
%\indent\indent\ The direct forcing term in Eq.~(1) is a simple harmonic drive, i.e., $F(t) = F\cos{\omega t}$ where $\omega$ is the drive frequency, while the feedback term can be expressed as $g(\dot{x}_1,\dot{x}_2) = G(s_1\dot{x}_1 + s_2\dot{x}_2)$ where $G$ is the complex feedback gain, and $s_{1,2} = 0,1$ indicate the exact combination of the feedback loop. For instance, when $s_1=1$, and $s_2=0$ we are applying a simple gain feedback on PMUT1, whereas if $s_1=0$, and $s_2=1$ we are feeding back the velocity of PMUT2 as a driving term to PMUT1.\\
\indent\indent\ An approximate solution to Eq.~(\ref{eqn:Eq1}) is obtained by applying the Rotating Wave Approximation (RWA) where the respective modal displacements are expressed as $x_{1,2} = \frac{1}{2}(A_{1,2}e^{i\omega t} + A^*_{1,2}e^{-i\omega t})$, where $A_{1,2}$ are the complex amplitudes of oscillations for PMUT1 and PMUT2, respectively. By expanding the terms in Eq.~(\ref{eqn:Eq1}) using the RWA and gathering the various first order $e^{\mathrm{i}\omega t}$ terms (see supplementary material for detailed derivation), we obtain the following governing slow-flow dynamics\\
\begin{equation}
\begin{split}
\mathrm{i} \begin{bmatrix} \dot{A}_1\\ \dot{A} _2 \end{bmatrix} = \frac{1}{2}\\[1ex]	\begin{bmatrix} (2\delta_1' +\frac{3\alpha}{4}\vert A_1 \vert ^2) -\mathrm{i}({\gamma_1'} + \frac{3\beta}{4}\vert A_1 \vert ^2) & \Gamma'\\ {\Gamma} & (2\delta_2-{\mathrm{i}}\gamma_2) \end{bmatrix}\begin{bmatrix} A_1\\A_2 \end{bmatrix} \\[1ex] + \frac{1}{2}\begin{bmatrix} {F}\\0 \end{bmatrix}
\end{split}
\label{eqn:Eq2}
\end{equation}\\
where $\delta_1$ and $\delta_2$ are the detunings between the oscillation frequency and the natural frequencies ($\omega_1$, and $\omega_2$) of PMUT1 and PMUT2, respectively (i.e., $\omega=\omega_{1, 2}(1+\delta_{1, 2})$), with $\delta_1' = \delta_1 + 0.5\times S_1\times\Re{(G)}$ while $\gamma_1' = \gamma_1+S_1\times\Im{(G)}$; $g(\dot{x}_1,\dot{x}_2) = S_1G\dot{x}_1 + S_2G\dot{x}_2$, where $G$ is the feedback gain, $\Gamma = \Gamma_{el} + \mathrm{i}\Gamma_{ac}$, and $\Gamma' = \Gamma + \mathrm{i}S_2\times G$. With $\gamma_{1, 2}$, $\delta_{1, 2}$, and $F$ being real valued, while $G$, $\Gamma_{el}$, and $\Gamma_{ac}$ take on complex values. $S_1$ and $S_2$ indicate which quantity is being fed back into PMUT1; If the motion of PMUT1 is fed back, then $S_1 = 1$ and $S_2 = 0$, while for a feedback of PMUT2's motion onto PMUT1, $S_1 = 0$ and $S_2 = 1$. Equation~(\ref{eqn:Eq2}) has been written in non-dimensional form where time is re-scaled as $t\rightarrow t\omega$, equally all system parameters are re-scaled as follows $\gamma_{1,2} \rightarrow \gamma_{1,2}/\omega$, $\Gamma_{el} \rightarrow \Gamma_{el}/\omega^2$, $\Gamma_{ac} \rightarrow \Gamma_{ac}/\omega$, $\beta \rightarrow \beta/\omega^2$, $F(t) \rightarrow F(t)/\omega^2$, and $g(\dot{x}_1,\dot{x}_2) \rightarrow g(\dot{x}_1,\dot{x}_2)/\omega^2$.\\%$g(\dot{x}_1,\dot{x}_2) \rightarrow g(\dot{x}_1,\dot{x}_2)/\omega^2$.\\
%\indent\indent\ Equation~(2) is capable of describing the full gamut of the dynamics of the pairs of coupled transducers with or without feedback by setting $s_1$, $s_2$, and $F$ to their respective values.\\
\indent\indent\ If only a forced steady-state response is sought, then we can set the left hand side of Eq.~(\ref{eqn:Eq2}) to zero and set $S_1=S_2=0$, which allows us to express the ratio of the amplitudes of the two PMUTs as (see Supplementary Material for derivation)
\begin{equation}
\frac{A_2}{A_1} = \frac{\Gamma}{2(\delta_2-\frac{\mathrm{i}}{2}\gamma_2)}
\label{eqn:Eq3}
\end{equation}\\
Equation~(\ref{eqn:Eq3}) indicates that the magnitude of vibration ratio of PMUT2 to PMUT1 would give a Lorentzian whose detuning ($\delta_2$) and linewidth ($\gamma_2$) are those of PMUT2, and whose amplitude directly depends on the coupling ($\Gamma$). This provides us with means to experimentally quantify the coupling parameter.\\
\begin{figure}%[th]%[p][th]
	\graphicspath{{Figures/}}
	\includegraphics[width=85mm]{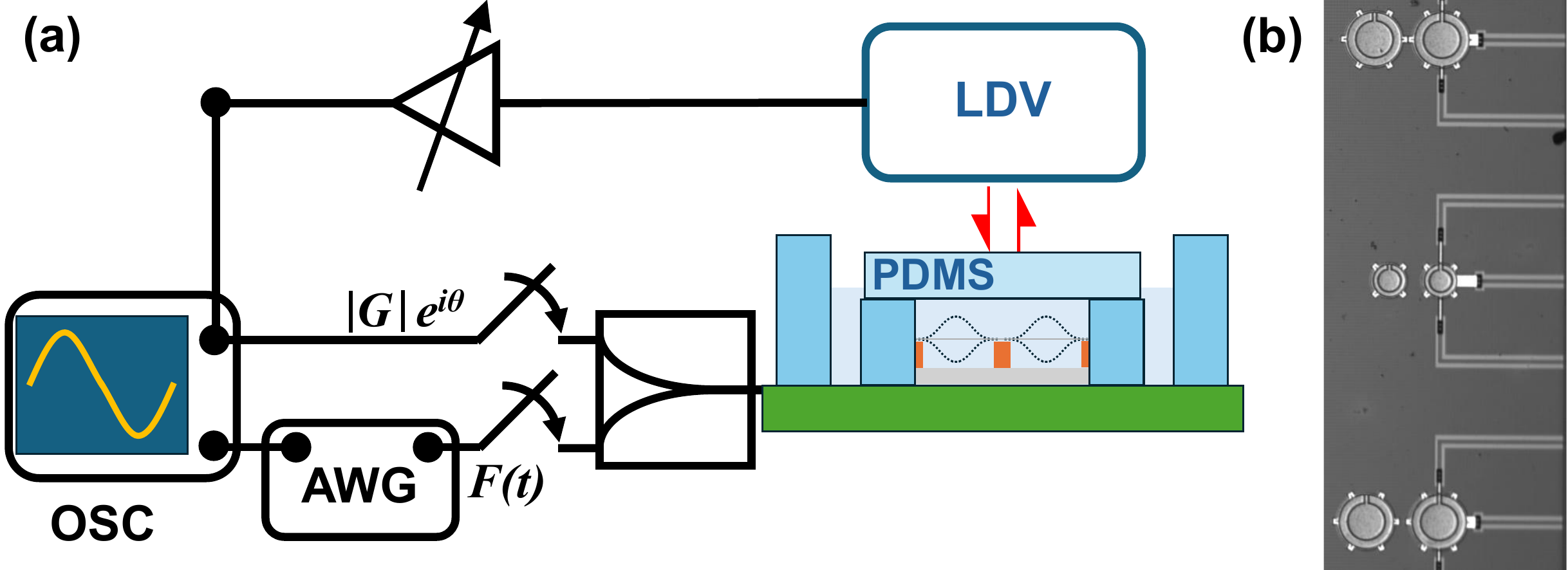}
	\caption{(a) Schematic representation of the measurement setup, showing the possibility to have either gain feedback loop or direct drive depending on which of the two switches is closed. Also shown is a schematic of the in-water measurement layout showing the PMUT sample (black and grey), the PMMA water containers (dark blue) and the PDMS acoustic absorber (light blue). (b)  Optical microscope picture of the experimental samples showing the driven and the un-driven MEMS devices, also visible are the alternating diameters per row (100 and 60 $\mu$m diameters).}
    \label{fig1}
\end{figure}
%\section{Experimental Results and Discussion}
\indent\indent\ Experimentally, the samples used in this work are fabricated by SilTerra Malaysia \cite{silterra} and consist of circularly shaped membranes formed mainly from a silicon nitride elastic layer, a Al$_{0.9}$Sc$_{0.1}$N piezoelectric layer which is sandwiched between two metal electrodes, and a cavity is etched under the stack, thus suspending the membranes. Details regarding the device stack and electrode shape optimization can be found in \cite{afshari2025biocompatible,chare2022electromechanical}.\\
\indent\indent\ In this work, the layout of the devices is optimized for the investigation of coupling between PMUT pairs. As such, pairs of devices with varying diameters are arranged in rows, with one device possessing electrical access for actuation while the second device in the pair is electrically unconnected, but otherwise identical. % The slight difference in the design of both devices in the pair leads to a slight discrepancy in the resonance frequencies, however, the impact of coupling remains clearly visible.
The pitch between devices is changed between each device pair in order to quantify the impact of separation on coupling. In all, four device diameters are investigated $40$, $60$, $80$, and $100$ $\mu$m with pitches ranging from 55 $\mu$m to 415 $\mu$m in steps of $20\mu$m (the exact pitch range depends on the diameter). The rows of a given diameter device pairs are interleaved with different diameter ones in order to increase the distance between equal diameter pairs and reduce any parasitic coupling.\\% Figure~\ref{fig1}(b) shows an optical microscope image of a sample of the PMUT pairs used in this study along with a schematic of the measurement setup.\\ 
\indent\indent\ The dies containing the devices are wire-bonded to a pcb and the bond pads are covered with epoxy to protect them during the in-water measurements. Two PMMA rings having 25 mm in diameter, 2 mm in thickness, and 12 mm in height are centered around the die and glued to the pcb and functions as a container that holds enough water to immerse the devices during measurements and enable the investigation of acoustic coupling in water under the vibrometer. For in-water measurements a 1 cm thick slab of PDMS is placed in contact with the water surface, its function is to attenuate acoustic waves reaching the surface and to reduce or eliminate acoustic reflections from the water-air interface \cite{weekers2023aln}. Figure~\ref{fig1}(a) shows a schematic representation of the measurement setup, while Fig.~\ref{fig1}(b) shows an optical microscope image of a sample of the PMUT pairs used in this study.\\
\indent\indent\ In the open-loop configuration ($S_1 = S_2 =0$ and $F\neq 0$) a computer controlled Arbitrary Waveform Generator (AWG, keysight 33600A) is used to drive the devices, the dies are placed under an MSA-500 Polytec laser Doppler vibrometer (LDV) whose output is sampled using a computer controlled oscilloscope (Tektronix TBS2000B). In case the feedback loop is active ($F=0$), the AWG is disconnected and the output from the LDV is amplified via a cascade of Transimpedance amplifier (Thor Labs AMP140) followed by a digitally controlled attenuator (Mini-Circuits ZX76-15R5A-PPS+) and a voltage amplifier (FEMTO	HVA-10M-60-F), these cascaded instruments are meant to provide a controllable gain. The amplified signal is then reintroduced to the device as a driving signal. The calibration of the amplifier gain can be found in Supplementary Material, it should be noted however, that the gain steps are not equally spaced due to the attenuator used. %By disconnecting the feedback loop, a direct drive arrangement is obtained, i.e., $S_1 = S_2 =0$ and $F\neq 0$. 
It should also be noted that the LDV offers the possibility to read-out only one location at a time, thus either the motion of PMUT1 or the motion of PMUT2 will be fed into only PMUT1, since it is the only one with electrical connections. Due to frequency limitations on the used amplifiers, the feedback-loop measurements are only applied to the 100 $\mu$m and 80 $\mu$m devices.\\ 
%\indent\indent\ A computer controlled Arbitrary Waveform Generator (AWG, keysight 33600A) is used to drive the devices, the dies are placed under an MSA-500 Polytec laser Doppler vibrometer (LDV) whose output is sampled using a computer controlled oscilloscope (Tektronix TBS2000B). In case the feedback loop is active, the output from the LDV is equally amplified via a cascade of Transimpedance amplifier (Thor Labs AMP140) followed by a digitally controlled attenuator (Mini-Circuits ZX76-15R5A-PPS+) and a voltage amplifier (FEMTO	HVA-10M-60-F), these cascaded instruments are meant to provide a controllable gain. The amplified signal is then reintroduced to the device as a driving signal. The calibration of the amplifier gain can be found in Supplementary Material. By disconnecting the feedback loop, a direct drive arrangement is obtained, i.e., $S_1 = S_2 =0$ and $F\neq 0$. It should be noted that since only PMUT1 is electrically connected, PMUT2 can not be measured during the application of the feedback loop as that means feeding the velocity of PMUT2 into PMUT1.\\ 
%\subsection{\label{sec:level2}Open-Loop Measurements}
\indent\indent\ As indicated by Eq.~(\ref{eqn:Eq3}) the ratio of PMUT2 to PMUT1 amplitudes results in a Lorentzian peak whose amplitude is directly proportional to the coupling, this is shown experimentally in-air (for 100 $\mu$m diameter and 135 $\mu$m pitch PMUTs) and in-water (for 100 $\mu$m diameter and 115 $\mu$m pitch PMUTs) in Fig.~\ref{fig2}(a) for open-loop measurements under the effect of a frequency sweep excitation. However, dividing the two somewhat equal amplitudes results in amplification of the noise as is clearly seen for the case of in-water measurement. To workaround the rapidly decreasing SNR of the amplitudes ratio, we approximate the response of the amplitudes ratio by first fitting independently the two peaks and then treating the ratio of the reconstructed peaks as a proxy to the ratio of the raw data.\\
\indent\indent\ Using such an approximation gives the coupling strength $\vert\Gamma\vert$ as a function of pitch for both in-air and in-water, as shown in logarithmic scale in Fig.~\ref{fig2}(b). In water, the observed coupling strength decreases roughly as a function of 1/pitch which is commensurate with the pressure profile of acoustic transducers \cite{anderson2018understanding}. %Interestingly COMSOL simulations for the 100 $\mu m$ devices exhibit very similar tendencies to those observed from the measurements, equally shown in Fig.\ref{fig2}(b), which tends to confirm that the approximation made above is a reasonable one, with the caveat that COMSOL overestimates the quality factor and thus the experimental quality factor was used to rescale the COMSOL data (for details regarding COMSOL simulation see Supplementary Materials). 
The fitted phase on the other hand, shows a weak dependence on the normalized pitch, i.e. pitch/$\lambda$, in water and a more rapidly changing one in air (except for the 40 $\mu$m devices).\\%, which we attribute to difficulties arising from the associated very short wavelength.\\%. Here COMSOL is used to simulate the amplitudes ratio for the in-water experiments for the 100 $\mu$m devices, while the mean experimental quality factor is used to obtain the coupling magnitude from the simulated amplitudes ratio (for details regarding COMSOL simulation see Supplementary Materials).\\
\begin{figure}%[th]%[p][th]
	\graphicspath{{Figures/}}
	\includegraphics[width=85mm]{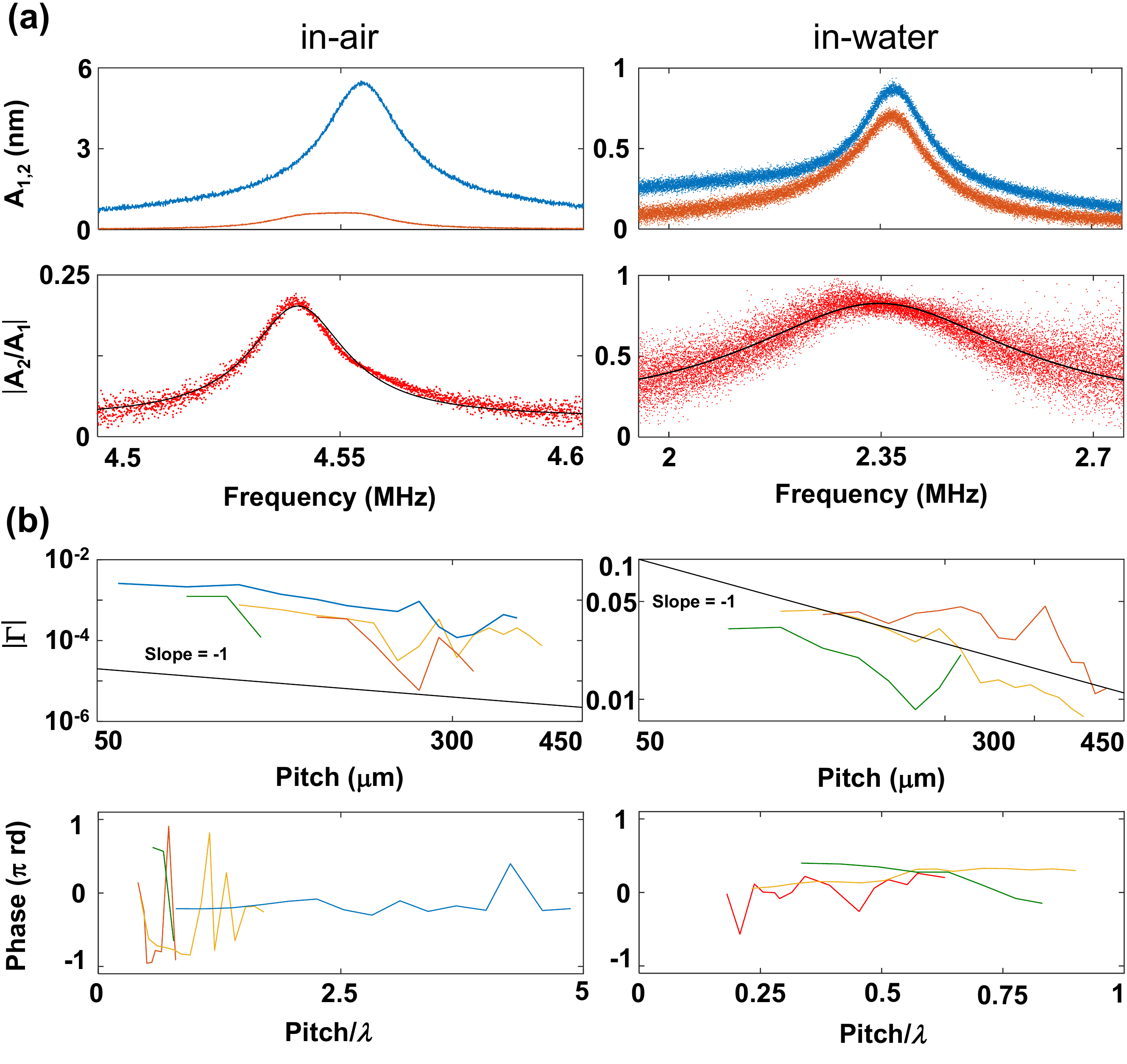}
	\caption{Coupling between two PMUTs. (a)  Resonance peaks (top) measured in air (left) and in water (right), and the ratio of the PMUT2 (red) to PMUT1 (blue) amplitudes (lower panels) showing that the ratio of the two indeed results in a Lorentzian peak, as well as demonstrating that the experimental noise is amplified when considering the ratio, especially for water measurements. (b) Dependence of the coupling strength on distance (top panels), in log-log scale, for the 100 $\mu$m (red), 80 $\mu$m (yellow), 60 $\mu$m (green), and 40 $\mu$m (blue), demonstrating a near 1/distance dependence both in air (left) and water (right). %COMSOL simulations for the 100 $\mu$m devices demonstrate similar tendencies (dashed red lines).
    Fitted coupling phase (lower panels) showing a slowly changing phase for all diameters in water, and a slowly changing phase for the 40 $\mu$m devices in air, but a rapidly changing one for the otehr diameters in air.}
    \label{fig2}
\end{figure}
\indent\indent\ Next, we apply the gain feedback loop in both configurations, since it is already clear from the results shown in Fig.~\ref{fig2} that the elastic coupling plays a negligible role in water, due to the low quality factor of the PMUTs in the latter and the presence of dominant acoustic coupling, as such we implement the feedback loop with only the water case and neglect the in-air case.\\
\indent\indent\ Figure~\ref{fig3} shows the effect of setting $F=0$ and applying the feedback either as $G\dot{x}_1$ (i.e., $S_1=1$, $S_2=0$), or as $G\dot{x}_2$ (i.e., $S_2=1$, $S_1=0$). Figure~\ref{fig3}(a) displays how the amplitude (linewidth) of the PMUTs increases (decreases) with the increasing loop gain leading to self-oscillation. The threshold of onset of self-oscillation can be readily identified, and is highlighted, in the pitch-gain parameter space as shown in Fig.~\ref{fig3}(b). Within the phase and gain limitations of the feedback loop used, the 100 $\mu$m devices are almost always oscillating when $S_1=1$, while in the case of the 80 $\mu$m devices, self-oscillation does not set in for $S_1=1$. On the other hand, the $S_2=1$ measurements show more interesting results with a strong dependence of the self-oscillation threshold on both pitch and gain.\\

 \begin{figure}[th]%[p][th]
	\graphicspath{{Figures/}}
	\includegraphics[width=85mm]{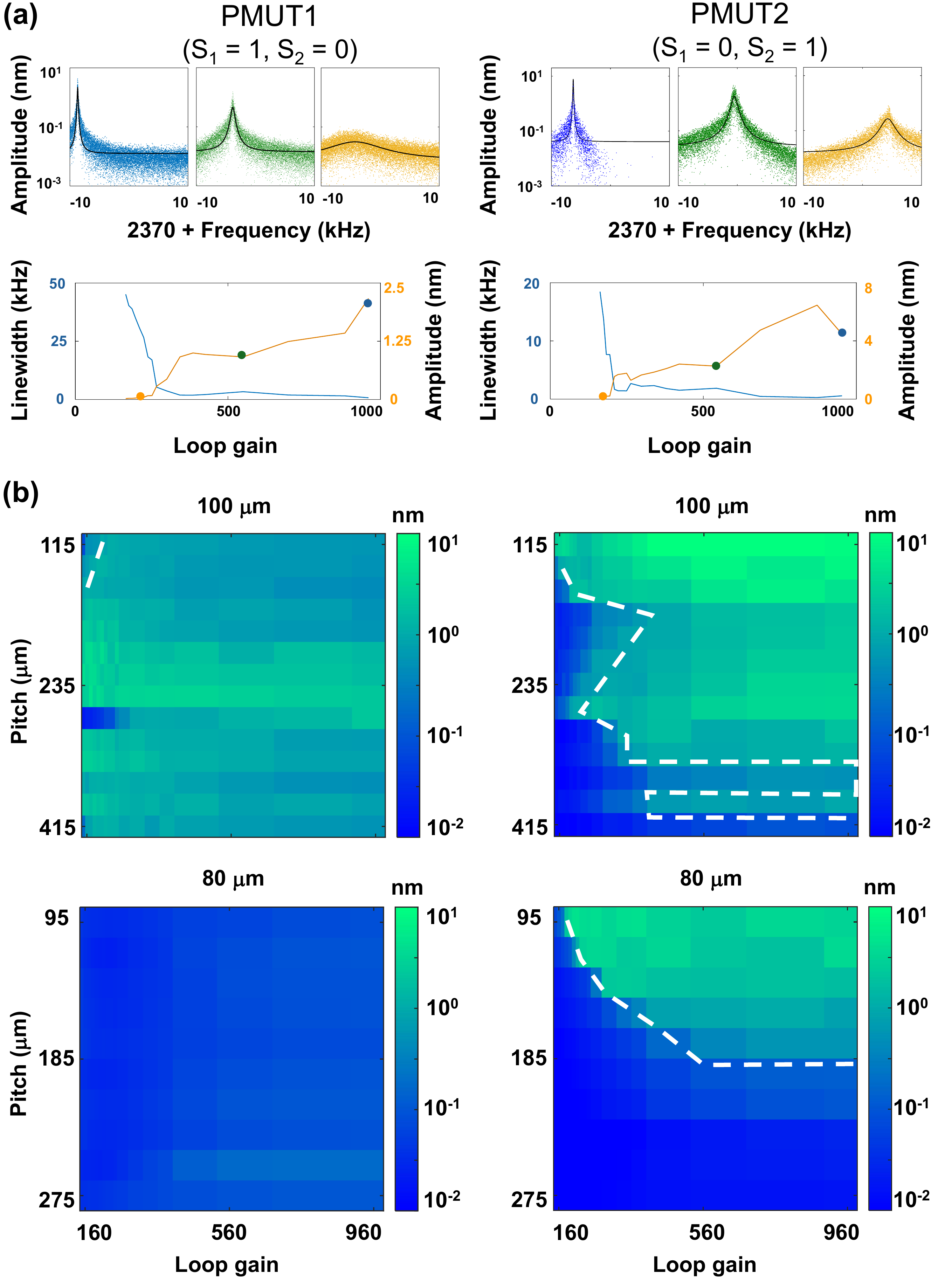}
	\caption{Self-oscillations and exceptional points. (a)  Effect of the feedback gain on resonance peaks for the 100 $\mu$m devices for 295 $\mu$m pitch (left) and 155 $\mu$m pitch (right) (top row), demonstrating the increase in quality factor, amplitude, and onset of self-oscillation for feedback with PMUT1 motion (left column) and PMUT2 motion (right column). Equally shown is the onset of oscillations as the gain is increased (bottom row). (b) 2D amplitude plots showing the onset of oscillations in the pitch-gain parameter space which is easily recognizable by the sudden increase in amplitude, the border is highlighted with a white dashed line for convenience, shown for the 100 $\mu$m devices (top) and the 80$\mu$m devices (bottom). For the used loop gain range, the feedback of PMUT1 motion shows constant (absent) oscillations for the (80) 100$\mu$m devices, therefore most of the coupling information is obtained from PMUT2 feedback data. The unequal gain step is visible in the plots.}
    \label{fig3}
\end{figure}
\begin{figure}%[p][th]
	\graphicspath{{Figures/}}
	\includegraphics[width=85mm]{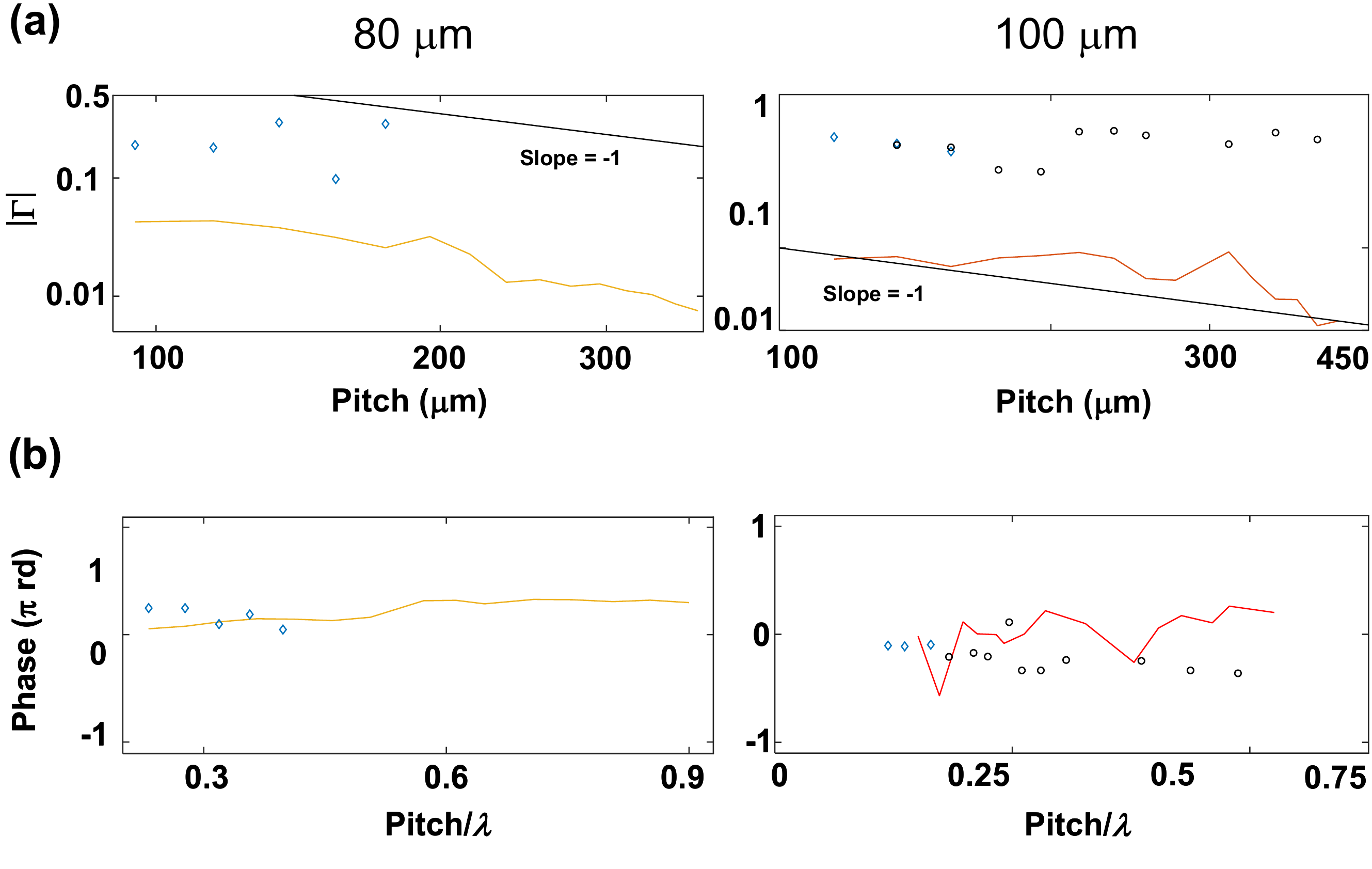}
	\caption{Coupling extracted from closed-loop measurements. (a)  Magnitude of coupling for the 100$\mu$m (right) and the 80$\mu$m (left) devices shown as discrete points are overlaid with values extracted from the open-loop measurements (continuous lines) %and COMSOL simulations (dashed lines). 
    For the 100 $\mu$m devices the Hopf bifurcation and the EP measurements show excellent agreement, the 80 $\mu$m devices have only Hopf-type data. (b) Extracted coupling phase showing a good agreement between Hopf and EP data, as well as the open-loop data.}
    \label{fig4}
\end{figure}

\indent\indent\ To extract the values of the coupling strength from the data in Fig.~\ref{fig3} it is crucial to have a proper understanding of the system in question. Indeed, whether the feedback is $G\dot{x}_1$ or $G\dot{x}_2$, represent two very different cases. To see why that is let us consider the eigenvalues ($E$) of the matrix in Eq.~\ref{eqn:Eq2}, dropping the nonlinear terms, gives $E_{\pm} = 0.5(T \pm (T^2-4\Delta)^{0.5})$, where $T$ and $\Delta$ are the trace and determinant of the matrix, respectively. When the feedback is that of $x_1$, the gain terms are in the matrix's diagonal as detailed in Eq.~\ref{eqn:Eq2}, leading to $T$ becoming purely real when $\gamma'=-\gamma$. If, in addition, the square root component of the eigenvalues is zero an exceptional point (EP) can form, having the coalesced eigenvalues $E_{\pm}= 2\delta + 0.5\Re{{(G)}}$ \cite{mao2023enhanced}. On the other hand if the feedback is $x_2$, then the feedback terms are in the anti-diagonal of the matrix (Eq.~\ref{eqn:Eq2}), thus the trace does not admit a non-complex value and as such no exceptional point can be formed. However, the system can still undergo self-oscillations via a Hopf bifurcation when the imaginary component of one of the eigenvalues crosses zero, this self-oscillation takes place without the eigenvalues coalescing, i.e, $E_+ \neq E_-$, and thus without the formation of an EP. Note that since $\Gamma$ is complex (unlike typically found in literature), the degenerate eigenvalues of the EP contain a gain-related term \cite{mao2023enhanced}.\\
\indent\indent\ The thresholds identified in Fig.~\ref{fig3}(b) are then fitted, using their respective models. The used models make the reasonable approximations (in-water) of $\gamma_1=\gamma_2=\gamma$, and $\delta_1=\delta_2=\delta$, which lead to the following relations to be fitted, $-\gamma \pm \Im{\sqrt{\Gamma(\Gamma+\mathrm{i}G)}}=0$, and $4\Gamma^2=-G^{*2}$ for the Hopf bifurcation and the EP cases, respectively. The extracted couplings are shown in Fig.~\ref{fig4} for the 100 $\mu$m and 80 $\mu$m devices. The data extracted from the feedback measurements are overlaid with those obtained from the open-loop measurements for visual comparison. The coupling values extracted from closed-loop measurements are larger than those obtained from the open-loop measurements (fig.~\ref{fig4}(a)), although the extracted phase seem to agree well (fig.~\ref{fig4}(b)). The extracted coupling shows a decaying trend with distance but the lack of sufficient number of data points impedes a good fit of this dependence. Equally interesting is that the coupling extracted from the Hopf and the EP-type measurements demonstrate very similar values.\\
\indent\indent\ In summary, this work investigated the effect of acoustic coupling on the behavior of PMUTs. In addition to an analytical model, three different experimental means for the quantification of coupling have been investigated, these rely respectively on open-loop and closed-loop measurements. Depending on which variable is used in the feedback loop either a PT symmetric EP or a Hopf bifurcation in the 2-degrees of freedom system are observed. The measured quantities are used to fit values for the coupling parameters which consistently show a distance dependence. The techniques used and coupling values extracted in this work can be used to better simulate and design 2-dimensional arrays of MEMS transducers.\\
\indent\indent\ See the supplementary material for a derivation of Eqs. (2) and (3), and calibration of feedback-loop gain.

\bibliography{apssamp}% Produces the bibliography via BibTeX.

\end{document}